\documentclass[
 reprint,
 amsmath,amssymb,
 aps,bm
]{revtex4-1}

\usepackage{natbib}
\usepackage{graphicx}
\usepackage{siunitx}
\usepackage{xcolor}  
\usepackage{textcomp}

\begin{document}

\title{Position-dependence of valley polarization and valley coherence of WS$_2$ monolayer flakes}

\author{Irina Komen}
\author{Sabrya E. van Heijst}
\author{Thomas Bauer}
\author{Sonia Conesa-Boj}
\author{L. Kuipers}
\email{L.Kuipers@tudelft.nl}

\affiliation{Quantum Nanoscience, Kavli Institute of Nanoscience, Delft University of Technology, The Netherlands}

\begin{abstract}
Chiral interaction between light and two-dimensional transition metal dichalcogenides (2D-TMDs) has recently drawn enormous scientific attention. The optical selection rules of these atomically thin semiconductors allow the attribution of a pseudospin to the TMDs' valleys, which can be coherently manipulated for information processing using polarized light. The interaction of TMDs with circularly and linearly polarized light creates the valley polarization and coherence, respectively. With a full Stokes polarization analysis of light emitted from WS$_2$ monolayer flakes, we conclusively confirm the existence of coherence between its valleys. We observe spatial heterogeneity in photoluminescence intensity, valley polarization and valley coherence. The discovery of an inverse proportional relationship between photoluminescence intensity and both valley polarization and coherence, reveals a correlation between the (non-)radiative decay rate and the valley hopping. The temperature dependence confirms the phononic nature of valley depolarization and decoherence, that may be caused by strain and the presence of defects.
\end{abstract}

\maketitle

\section{Introduction}

Two-dimensional transition metal dichalcogenides (2D-TMDs) have been widely investigated in the past years. These atomically thin semiconductor analogues of graphene (e.g. MoS$_2$, WS$_2$, MoSe$_2$, WSe$_2$, etc.) exhibit distinct physical phenomena that hold the promise of applications in opto-electronics \cite{Zhang_TMDCtransistor_science_2014, Wang_TMDCelectronics_NatNano_2012, Mak_TMDlightValley_NatPhot_2018}. In the monolayer limit, 2D TMDs possess a direct bandgap with transition frequencies in the (near) visible spectral range \cite{Mak_MoS2monofirst_PhysRevLett_2010, Splendiani_MoS2luminescence_NanoLett_2010}. Due to their high binding energy, electron-hole pairs form stable excitons even at room-temperature \cite{Chernikov_excitonBinding_PRL_2014, He_excitonBinding_PRL_2014}. The selection rules for optical transitions in the non-degenerate TMDs valleys allow the attribution of a pseudospin and selectively addressing each valley separately by using circularly polarized light of opposite handedness \cite{Zeng_TMDCvalleyPolarization_NatNano_2012, Mak_TMDCvalleyPolarization_NatNano_2012, Cao_TMDCcircular_NatCom_2012, Xu_TMDCspins_NatPhys_2014, Aivazian_TMDCmagneticValley_NatPhys_2015, Zhu_WS2bilayerValleyPolarization_PNAS_2014, Jones_TMDCvalleycoherence_NatNano_2013}. Important steps have been made in the search for valleytronics applications, where the valley pseudospin is used as a quantum number to encode and process information \cite{Mak_TMDCvalleyHall_science_2014, Schaibley_valleytronics_NatRev_2016, Irina_2020, SuHyun_2018}. 

2D TMDs layers can be fabricated by chemical vapor deposition (CVD) \cite{Song_CVDgrownWS2_ACSNano_2013, Zhang_CVDgrownWS2_ACSNano_2013, Cong_CVDgrownWS2_AdvOptMat_2014, Orofeo_CVDgrownWS2_APL_2014, Thangaraja_WS2crystals_MatLett_2015, Liu_CVDgrownWS2_NanoscResLett_2017}. Interestingly, the photoluminescence (PL) intensity of CVD grown single layered TMDs (monolayers) is non-uniform across the flakes \cite{Gutierrez_PLpattern_NanoLett_2013, Peimyoo_growthPL_ACSNano_2013, Cong_growthPL_AOM_2020, Kim_PLpatternBiexciton_ACSNano_2016, Liu_PLpattern_NanoLetters_2016, Feng_PLpatternStrain_ACS_2017, Creary_PLpatternValley_ACSNano_2017, Rosenberger_PLpatternDefect_ACSNano_2018}. Many explanations have been proposed to explain the non-uniformity in the PL intensity, including growth conditions \cite{Peimyoo_growthPL_ACSNano_2013, Cong_growthPL_AOM_2020}, differences in defect density  \cite{Peimyoo_growthPL_ACSNano_2013, Rosenberger_PLpatternDefect_ACSNano_2018, Kastl_PLpatternDefect_ACSNano_2019}, variations in chemical composition \cite{Liu_PLpattern_NanoLetters_2016, Gutierrez_PLpattern_NanoLett_2013}, grain boundaries \cite{Kim_PLpatternBiexciton_ACSNano_2016} and variations in strain \cite{Feng_PLpatternStrain_ACS_2017}. The chiral optical selection rules and the resulting valley pseudospin are a fundamental aspect of the electronic bandstructure of TMDs materials, and a key feature for potential applications. Therefore, investigating the interaction of the TMDs valleys with polarized light will clarify the processes that result in the inhomogeneous PL intensity. 

\begin{figure*}[htp]
\centering
\includegraphics[width = 0.9\linewidth] {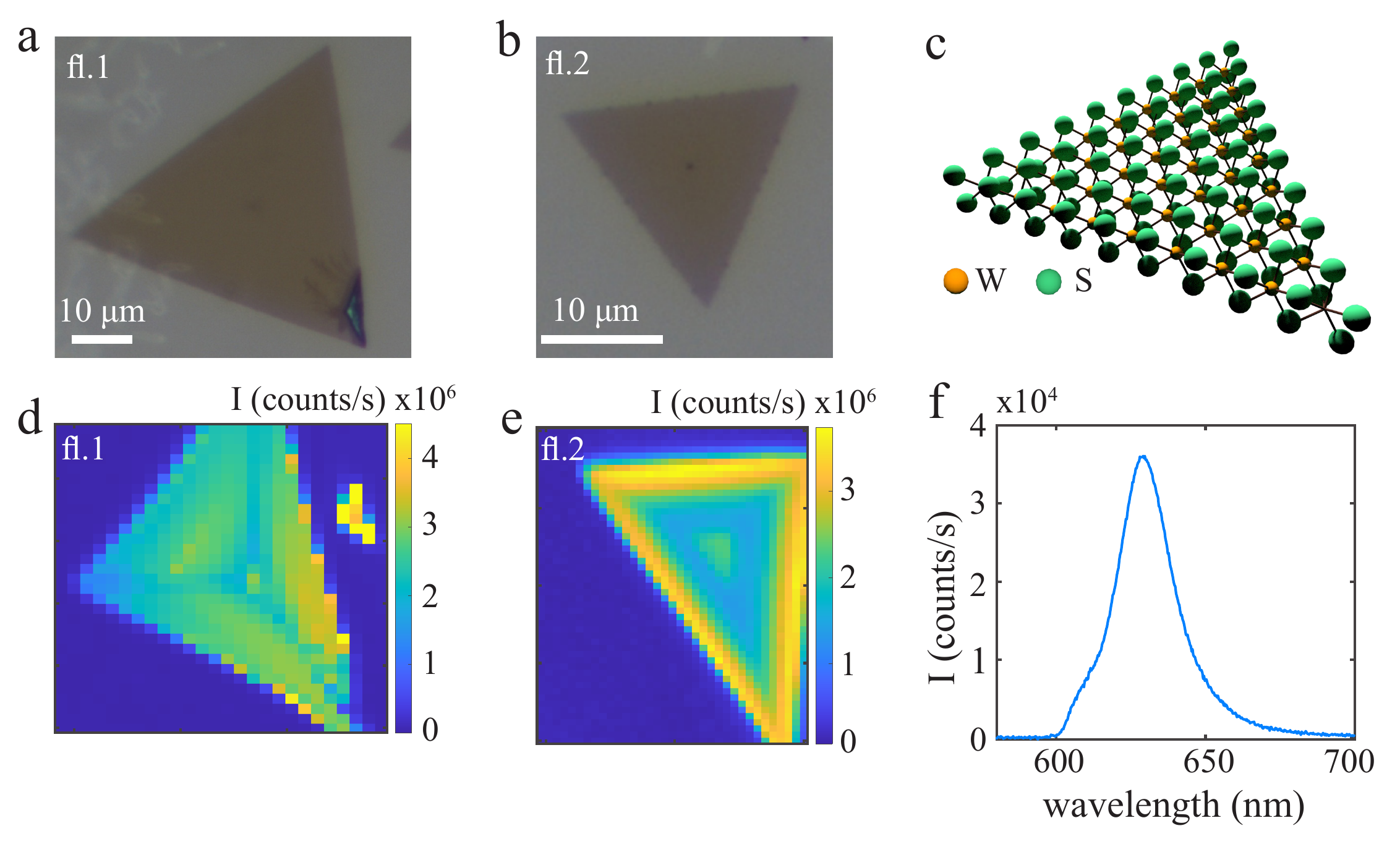}
\caption{\textbf{WS$_2$ monolayer flakes} \\
\textbf{a-b.} Optical (wide-field) images of CVD grown WS$_2$ monolayer flakes. The flake sizes are 55 and 20$\mu$m (see the scalebar). \textbf{c.} Schematic representation of a triangular WS$_2$ monolayer flake, depicting the atomic hexagonal structure that induces the growth in a triangular geometry. \textbf{d-e.} Raster scan of the non-uniform photoluminescence (PL) intensity over two monolayer flakes. Flake 1 exhibits lower intensity along the triangle medians, dividing the flakes into three regions with higher intensity. Flake 2 exhibit higher intensity along the edges. Note that the x and y axis in the raster scan are slightly skewed due to experimental constraints (see Supplementary Materials, Section III) \textbf{f.} Representative PL spectrum from WS$_2$ monolayer flake 1, acquired at room temperature. The high intensity and absence of indirect bandgap emission demonstrates that this is a monolayer and not a multilayer.}
\label{PLpattern}
\end{figure*}

Two relevant keywords in the polarization behaviour of the TMDs' photoluminescence are valley polarization and valley coherence. For valleytronics applications, it is paramount to generate a large asymmetric occupation of electrons in the two inequivalent valleys, i.e., a high valley polarization. Moreover, any presence of coherence between the TMDs valleys holds the intriguing promise of quantum manipulation of the valley index \cite{Jones_TMDCvalleycoherence_NatNano_2013, Zhu_WS2bilayerValleyPolarization_PNAS_2014, Wang_PolarizationWSe2magnets_PRL_2016, Schmidt_CoherenceWS2magnets_PRL_2016, Cadiz_polarizationMoS2magnets_PRX_2017}. To date, a limited number of measurements of linear polarization contrast have been performed and the findings attributed to valley quantum coherence \cite{Zhu_WS2bilayerValleyPolarization_PNAS_2014, Ye_valleyCoherenceTime_WSe2_NatPhys_2017, Hao_valleyCoherenceTime_NatPhys_2016}. Zhu et al measured a 4\% linear polarization contrast on a monolayer WS$_2$ at 10K \cite{Zhu_WS2bilayerValleyPolarization_PNAS_2014}. This suggests the presence of a quantum coherence between the two WS$_2$ valleys. However, a linear polarization contrast measurement does not allow to distinguish between unpolarized and circularly polarized components. Here, a full Stokes polarization analysis will add the missing information, e.g., it yields the full polarization state of the emitted light on the Poincar\'e sphere \cite{Lorchat_MuellerWS2graphene_ACSPhot_2018}. Therefore, a full Stokes polarization analysis of the TMDs emission would conclusively prove the existence of valley coherence. 

Here, we investigate the position-dependent photoluminescence intensity, valley polarization and valley coherence of CVD-grown, triangular WS$_2$ monolayer flakes. By performing a full Stokes polarization analysis on the WS$_2$ emission, we confirm the existence of coherence between the two WS$_2$ valleys. Surprisingly, we observe that high valley-polarization and high valley-coherence regions exhibit lower PL intensity. This anti-correlation between on the one hand intensity and on the other valley-polarization and -coherence is confirmed in multiple monolayer flakes, both at room temperature and cryogenic temperatures. The temperature-dependence reveals that phonons are the main source of valley depolarization and valley decoherence. The polarization behaviour of the emission of WS$_2$ monolayer flakes provides a tool to understand the interaction, hopping and coherence of electrons in the WS$_2$ valleys. 

\section{Results and Discussion}

\subsection{Photoluminescence intensity}

In Figure \ref{PLpattern}a-b we present optical microscopy images of WS$_2$ monolayer flakes with lateral dimensions of \SI{55}{\mu m} and \SI{20}{\mu m}, numbered fl.1 and fl.2. The triangular growth morphology of the WS$_2$ flakes is induced by the underlying hexagonal atomic structure, which is schematically represented in Fig.\ref{PLpattern}c. \mbox{Figure \ref{PLpattern}f} depicts a typical photoluminescence spectrum of a WS$_2$ flake. The high intensity and absence of indirect bandgap emission at higher wavelengths indicates that this is a monolayer rather than a multilayer flake. 

In Figure \ref{PLpattern}d-e we present the position-dependent PL intensity of the monolayer flakes (determined by integrating the area under every PL spectrum). Clearly, the photoluminescence intensity is non-uniform, exhibiting regions of higher and lower intensity. The pattern of the PL intensity seems to divide fl.1 into three smaller triangles with a higher intensity in the centre, and a low PL intensity along the flake medians (Fig.\ref{PLpattern}d). Fl.2 exhibits higher PL intensity at the edges (Fig.\ref{PLpattern}e). The position-dependent PL of other monolayer flakes exhibits similar intensity patterns, e.g., either a lower intensity along the flake medians or a higher intensity at the edges (see Supplementary Materials Fig.S1 for other examples). The Scanning Electron Microscopy (SEM) images of the monolayer flakes do not reveal any grain boundaries or other structural reasons that would explain the spatial non-uniformity in the PL intensity (see Supplementary Materials Fig.S1 for the SEM images). 

Comparing the PL spectra originating from the very middle to spectra from the rest of the flake, we find significant spectral differences (see Supplementary Materials, Fig.S1). All other spectra, both obtained from higher and lower intensity regions, do not exhibit differences in shape or spectral position, but only in intensity. The spectra from the very middle of fl.1 and fl.2 however are significantly red shifted. Based on the spectral position of this shifted peak, we attribute it to the presence of local structural defects. Not all the monolayer flakes exhibit these defect-related spectra (see Supplementary Materials Fig.S1 for examples). Structural characterization on the atomic level cannot be performed on the investigated monolayer flakes, as the relatively thick silicon substrate is not transparent to Transmission Electron Microscopy. 

Similar position-dependent variations in the PL intensity of CVD grown TMDs monolayer flakes have been reported before \cite{Gutierrez_PLpattern_NanoLett_2013, Peimyoo_growthPL_ACSNano_2013, Cong_growthPL_AOM_2020, Kim_PLpatternBiexciton_ACSNano_2016, Liu_PLpattern_NanoLetters_2016, Feng_PLpatternStrain_ACS_2017, Creary_PLpatternValley_ACSNano_2017, Rosenberger_PLpatternDefect_ACSNano_2018}. A correlation was observed between the size of the monolayer flakes and PL intensity pattern \cite{Liu_PLpattern_NanoLetters_2016}. The intensity pattern of fl.1, with the low intensity triangle medians, indeed matches the reported patterns for monolayer flakes of size \SI{55}{\mu m}, and the intensity pattern of fl.2, with the high intensity along the edges, matches the reported pattern for monolayer flakes of size \SI{20}{\mu m}. 

\begin{figure*}[htp]
\centering
\includegraphics[width = \linewidth] {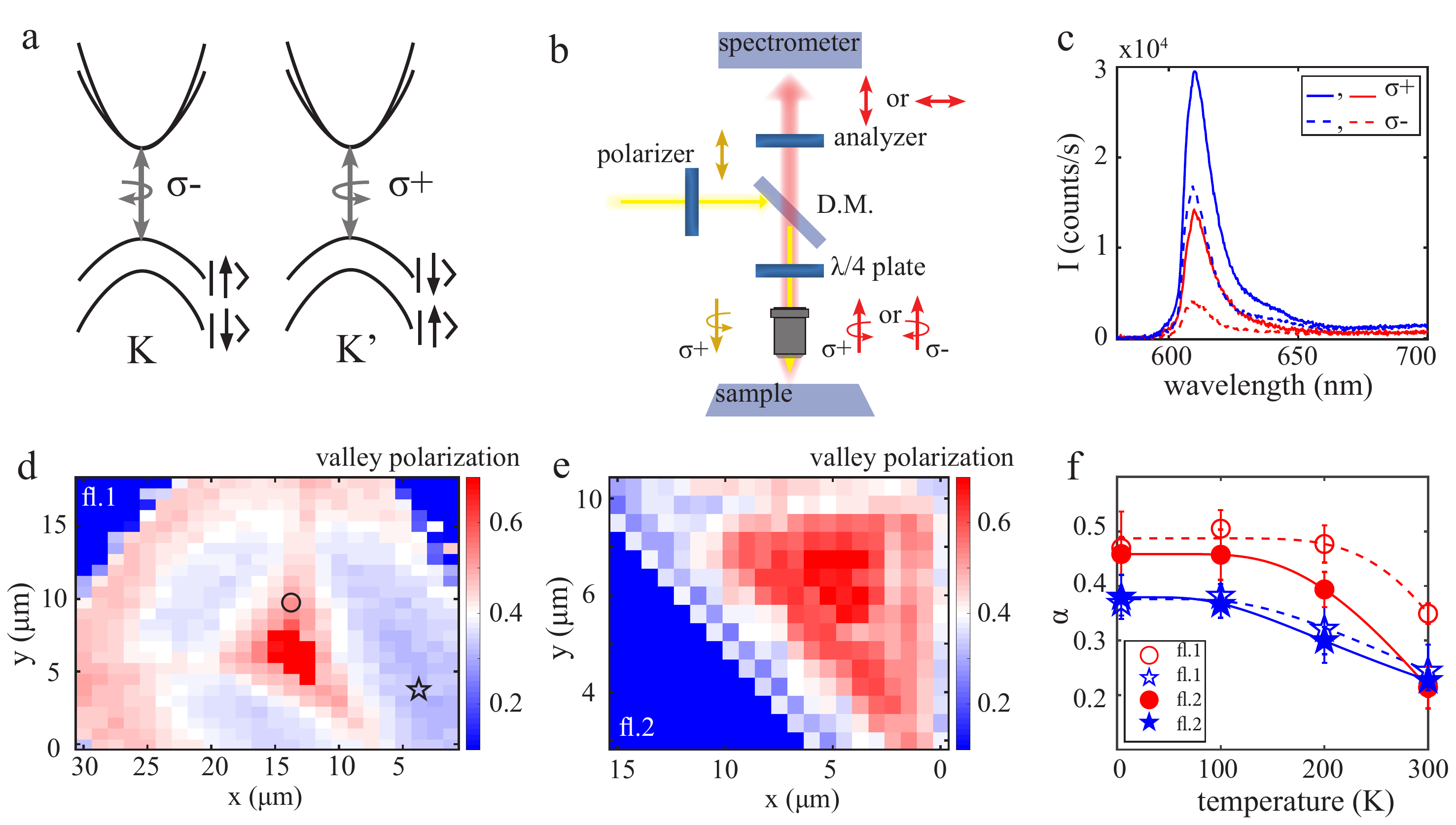}
\caption{\textbf{Position-dependent valley polarization} \\
\textbf{a.} Valley-dependent optical transitions of WS$_2$, where the two nonequivalent valleys at the K and K' points can be optically addressed with a specific circular polarization. \textbf{b.} Schematic of our set-up, where the circularly polarized excitation light is focused on the sample through an objective lens. The emission is collected through the same objective lens and quarter-wave plate, and directed to a spectrometer through an analyzer. \textbf{c.} PL spectra on fl.1 taken at a temperature of \SI{4}{K}, on a high intensity region (blue) and a low intensity region (red) (star and circle in \textbf{d}). The WS$_2$ is excited with $\sigma_+$ light, and from the difference in the $\sigma_+$ and $\sigma_-$ emission we determine the degree of valley polarization. \textbf{d,e.} Degree of valley polarization of fl.1 and fl.2. A clear distinction can be seen between regions with high valley polarization (in red) and low valley polarization (in blue). Note that the regions with lower intensity exhibit higher valley polarization than the lower intensity regions on these flakes (compare with Fig.\ref{PLpattern}d,e). \textbf{f.} Temperature dependence of the average valley polarization of the low and high valley polarization regions (blue stars vs. red circles) of fl.1 and fl.2 (open vs. filled markers). Note that the bars do not indicate errors, but rather the standard deviation of the distribution of the measured valley polarization obtained over the specific region of the flake. The temperature dependence is fitted with a Boltzmann distribution, assuming valley depolarization is caused mainly by phonons.}
\label{valleyPol}
\end{figure*}

\subsection{Valley polarization}

The observed spatial heterogeneities in the PL intensity in Fig.\ref{PLpattern}d,e raise the question about the potential spatial variations in the valley polarization. A correlation was previously reported between the PL intensity of monolayer flakes and the valley polarization \cite{Creary_PLpatternValley_ACSNano_2017}. We therefore determine the valley polarization as a function of position for the triangular WS$_2$ monolayers at both room- and cryogenic temperatures. The WS$_2$ valleys with their valley-dependent optical selection rules are schematically depicted in Fig.\ref{valleyPol}a. Valley polarization is measured by addressing one of the TMDs valleys with circularly polarized incident light and determining the degree of circular polarization of the resulting PL. \mbox{Figure \ref{valleyPol}b} depicts a schematic of the employed experimental set-up. The excitation light (\SI{595}{nm} wavelength) is given a specific circular polarization by a combination of a polarizer and quarter-wave plate. The circularly polarized excitation light is focused on the sample (see Methods for experimental details), and the emission is collected in reflection through the same objective and projected onto a linear polarization state by the same quarter-wave plate used to polarize the excitation beam. Comparing the vertical and horizontal components of the resulting emission light using a polarization analyzer and a spectrometer, we determine the relative contributions of both circular polarization components. 

\mbox{Figure \ref{valleyPol}c} depicts photoluminescence spectra from the high-intensity regions (blue spectra) and low-intensity regions (red spectra) of fl.1 (obtained at \SI{4}{K}). We excite the WS$_2$ with $\sigma$+ light, and from the difference in the $\sigma$+ (solid line) and $\sigma$- (dashed line) emission, we determine the degree of valley polarization $\alpha = \frac{I_+ - I_-}{I_+ + I_-}$ (with I$_+$ and I$_-$ determined by integrating the area under the $\sigma$+ and $\sigma$- PL spectra). For fl.1 (see Fig.\ref{PLpattern}d) the low-intensity spectra on the medians (in red) yield a valley polarization of 0.46+-0.02, whereas on the high-intensity spectra of the monolayer (in blue) the valley polarization is significantly lower, $\alpha$ = 0.37+-0.02. \mbox{Figures \ref{valleyPol}d-e} present the valley polarization of (the central region of) fl.1 and fl.2 at \SI{4}{K}. The valley polarization of fl.1 follows a similar pattern as the PL intensity, dividing the flake into three smaller triangles with lower valley polarization (blue in the Fig.\ref{valleyPol}d), and exhibiting a higher valley polarization along the flake medians (red in Fig.\ref{valleyPol}d). When comparing the valley polarization in Fig.\ref{valleyPol}d with the PL intensity in Fig.\ref{PLpattern}d, it becomes apparent that the WS$_2$ flake exhibits high valley polarization in the regions of low PL intensity (along the flake medians), and low valley polarization in the regions of high PL intensity. The same inverse relationship between intensity and valley polarization is observed in fl.2 (compare Fig.\ref{PLpattern}e), where the lower intensity region in the middle of the monolayer flake exhibits a higher valley polarization (red in Fig.\ref{valleyPol}e) than the high intensity edges (blue in Fig.\ref{valleyPol}e). We conclude that the valley polarization and the PL intensity are inverse proportional. It is interesting to note that the valley polarization of fl.3 and fl.4, which do not exhibit defect-related spectra, is more homogeneous across these flakes (see Supplementary Materials Fig.S5). Nevertheless, fl.3 and 4 between them exhibit the same inverse relationship between intensity and valley polarization as fl.1 and 2, where the correlation was observed within a single flake. 
In order to understand the processes that determine the degree of valley polarization better, we investigate its temperature dependence. The spatial variations and the specific patterns of high and low valley polarization, as presented in Fig.\ref{valleyPol}d-e, are present both at room- and at cryogenic temperatures (see Supplementary Materials Fig.S3 for valley polarization maps at different temperatures). Figure \ref{valleyPol}f depicts the temperature dependence of the average valley polarization of the high valley polarization (red circles) and low valley polarization (blue stars) regions of fl.1 and fl.2 (open vs. filled markers, compare blue and red regions in Fig.\ref{valleyPol}d-e). In all cases, the valley polarization decreases with increasing temperature. Following Zhu et al \cite{Zhu_WS2bilayerValleyPolarization_PNAS_2014}, we fit the temperature-dependent valley polarization by a Boltzmann distribution, assuming that phonons are the main source for valley depolarization. From the good quality of fit we conclude that valley depolarization is mainly caused by phonons. 

\begin{figure}[hbp]
\centering
\includegraphics[width = \linewidth] {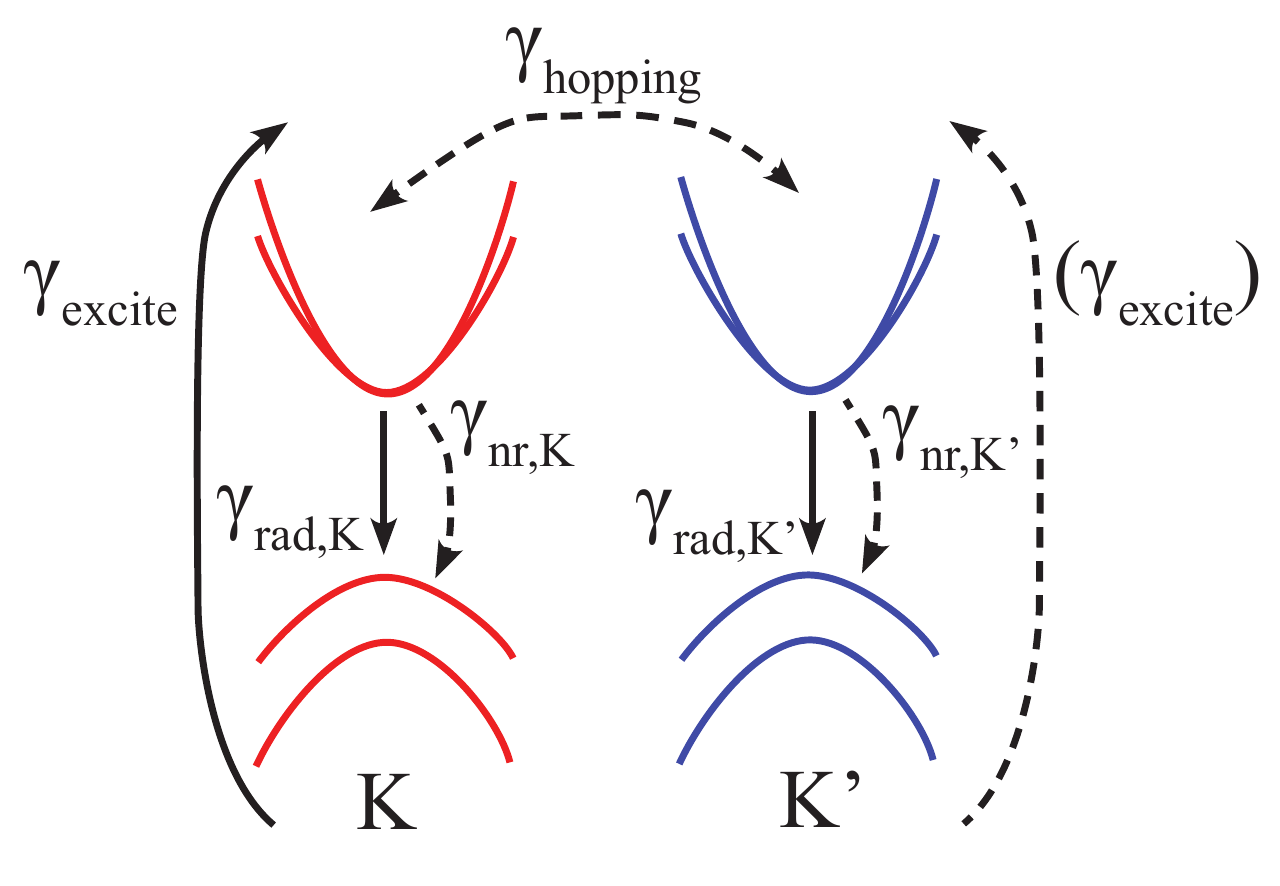}
\caption{\textbf{Excitation and decay processes in WS$_2$ valleys} \\
Schematic of the processes in the WS$_2$ valleys with their different rates. The radiative vs. non-radiative decay rates ($\gamma_{rad}$ and $\gamma_{nr}$) influence the photoluminescence intensity, and the valley hopping $\gamma_{hopping}$ reduces the degree of valley polarization and valley coherence.
}
\label{scheme_valleys}
\end{figure}

Figure \ref{scheme_valleys} schematically presents all the relevant processes in the WS$_2$ valleys. After the selective excitation of one of the WS$_2$ valleys ($\gamma_{excite}$ in Fig.\ref{scheme_valleys}), there are several possible scenarios: the electron can hop to the other valley ($\gamma_{hopping}$), the electron can decay radiatively ($\gamma_{rad,K}$) or non-radiatively ($\gamma_{nr,K}$). On the one hand, a fast radiative decay and a slow non-radiative decay results in a higher PL intensity. On the other hand, the valley hopping rate reduces the valley polarization. Differences in intensity and valley polarization on CVD grown flakes have been attributed before to differences in radiative and non-radiative lifetime components \cite{Creary_PLpatternValley_ACSNano_2017}. We observe an inverse relationship between PL intensity and valley polarization, which implies a correlation between the valley hopping rate and the (non-)radiative decay rate. More specifically, we observe a high valley polarization at regions of low PL intensity. We conclude that in these regions, the valley hopping rate is relatively slow with respect to the rest of the flake and the non-radiative decay rate is relatively fast. At regions of high PL intensity we observe a low valley polarization. In these regions therefore the valley hopping rate is relatively fast and the non-radiative decay rate is relatively slow. 

\begin{figure*}[htp]
\centering
\includegraphics[width = \linewidth] {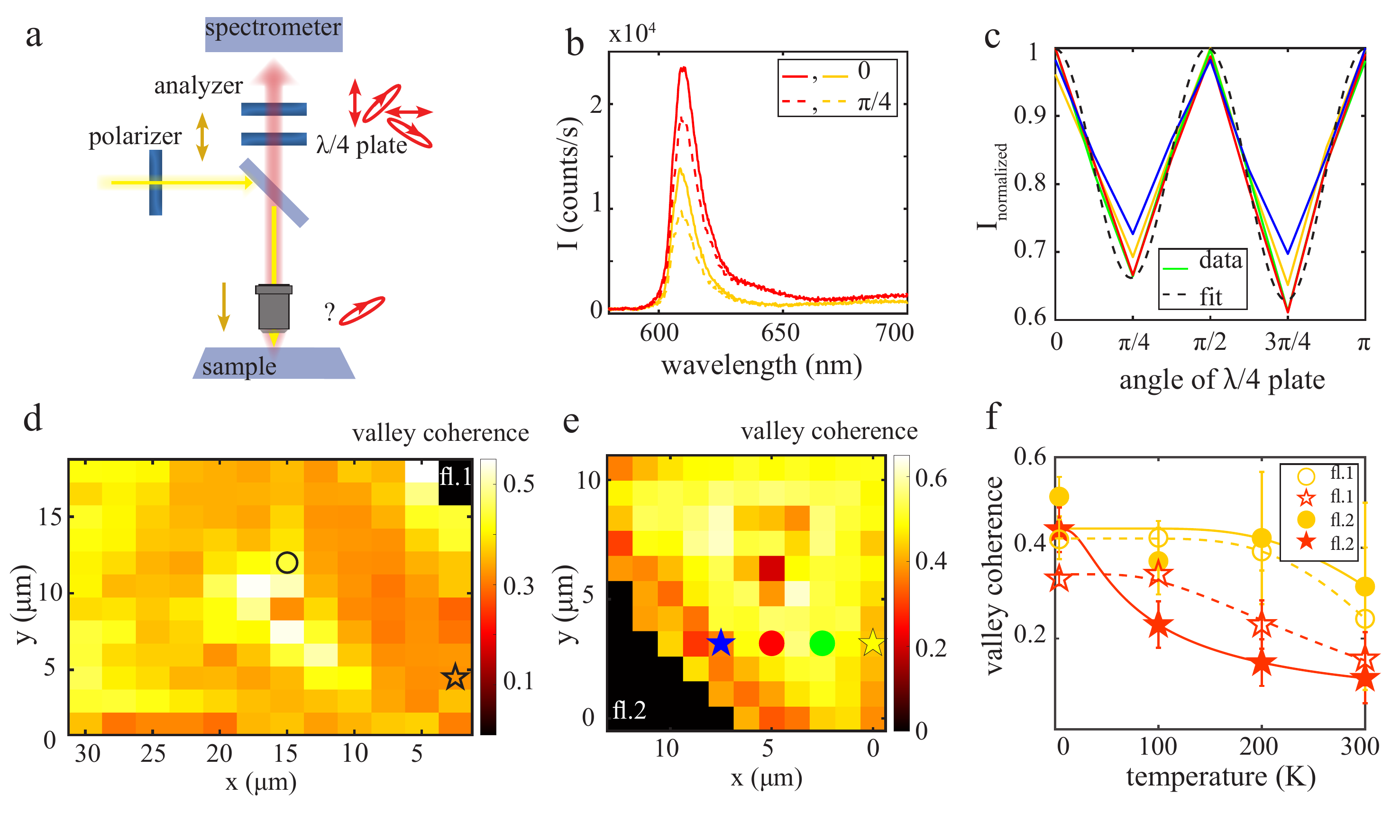}
\caption{\textbf{Position-dependent valley coherence} \\
\textbf{a.} Schematic of the set-up, where linearly polarized excitation light is focused on the sample. The emission is directed to a spectrometer, passing through a quarter-wave plate (rotated from 0 to $\pi$ in steps of $\pi/8$) and analyzing polarizer (fixed along 0). \textbf{b.} PL spectra on fl.1 at \SI{4}{K}, on a high intensity region (red spectra), and a low intensity region (yellow spectra) (star/circle in \textbf{d}). \textbf{c.} Normalized intensity as a function of quarter-wave plate angle, at high and low intensity regions on fl.2 (stars/circles in \textbf{e}). The curves at high intensity (blue and yellow) exhibit a dip of only 0.75 at an angle of $\pi/4$ and at low intensity (red and green) a dip of 0.65. The Stokes parameters and thus the polarization coherence can be extracted via Fourier analysis of these curves. \textbf{d,e} Valley coherence of the middle regions of fl.1 and fl.2. Note that the regions with lower intensity have a higher valley coherence than the high intensity regions. \textbf{f.} Temperature dependence of the average valley coherence of the low and high valley coherence regions (red stars vs. yellow circles) of fl.1 and fl.2 (open vs. filled markers). (Note that the bars do not indicate errors, but rather the standard deviation of the distribution of the measured valley polarization obtained over the specific region of the flake). The temperature dependence of the valley coherence can be fitted with a Boltzmann distribution, assuming valley decoherence is caused mainly by phonons.}
\label{coherence}
\end{figure*}

\subsection{Valley coherence}

The valley coherence reflects the interaction and phase relation of both valleys. Valley coherence is measured by exciting both valleys equally using linearly polarized light, which is a coherent superposition of left-handed and right-handed circularly polarized light. If we suppose the presence of valley coherence in a TMDs material, the coherence of the excitation photons is transferred to the valleys, so that the pseudospins associated to the two TMDs valleys will form a superposition state. In most semiconductors, phase information is lost in the photoluminescence process \cite{dragoman_opticsSolids_2002}, therefore the random phase of the emission of different valleys or atoms always results in unpolarized light. However, if the pseudospins associated with the two TMDs valleys retain their coherence perfectly, the emitted photons exhibit a coherent superposition of left-handed and right-handed circularly polarized light again, e.g. the emission will be linearly polarized. 

By performing a full Stokes analysis \cite{Schaefer_stokesPol_APL_2007, collett_quarterwaveplate_1993, collett_quarterwaveplate_2005} on the emission of the flakes, we determine the exact polarization state of the emitted light and therefore gain access to the valley coherence. With this method, we measure the PL intensity as a function of the angle of a rotating quarter-wave plate and retrive the Stokes parameters using Fourier analysis \cite{Schaefer_stokesPol_APL_2007, collett_quarterwaveplate_1993, collett_quarterwaveplate_2005} (see also Supplementary Materials Section II). The Stokes parameters both provide information on the polarization state of the PL, e.g., whether the PL has components that are circularly or linearly polarized, as well as on the amount of unpolarized vs. polarized light. 

\mbox{Figure \ref{coherence}a} presents a schematic of our set-up, where linearly polarized excitation light (\SI{595}{nm} wavelength) is focused on the sample, after which the emission is collected and directed towards the spectrometer (see Methods). The emission passes through a quarter-wave plate, that is rotated from 0 to $\pi$ in steps of $\pi/8$, and a fixed analyzer (oriented along $\theta$=0). \mbox{Figure \ref{coherence}b} presents PL spectra of fl.1 acquired using the rotating quarter-wave plate method. The spectra are from a low-intensity region (in yellow) and a high-intensity region (in red) (circle vs star in Fig.\ref{coherence}d). The spectra with a 0 waveplate angle (solid line) have a higher intensity than the ones with a $\pi/4$ angle (dashed line). Spectra as a function of quarter-wave plate angle and position on the flake are measured on both fl.1 and fl.2. \mbox{Figure \ref{coherence}c} depicts the derived PL intensity on different regions of fl.2 as a function of the quarter-wave plate angle (see circles/stars in Fig.\ref{coherence}e). The derived PL intensity exhibits minima at $\pi/4$ and $3\pi/4$ of around 0.75 at high PL intensity (blue and yellow curves) and around 0.60 at low PL intensity (red and green curves). We attribute the small intensity differences between $\pi/4$ and $3\pi/4$ to slight imperfections in the optical alignment and polarization elements (see Supplementary Materials Section III for details). The depth of the intensity minima at $\pi/4$ and $3\pi/4$ provides information about the polarization state of the WS$_2$ emission. If the phase of the WS$_2$ emission would be fully incoherent, the emission intensity would be constant for all quarter-wave plate angles, whereas for linearly polarized emission we would expect a dip of 0.50 at $\pi/4$ and $3\pi/4$. The measured 0.60-0.75 dips in the PL intensity therefore confirm that the WS$_2$ emission exhibits partial phase coherence. 

To derive the value of the position-dependent valley coherence, we perform a Fourier analysis of the measured intensity curves as depicted in Fig.\ref{coherence}c, and first derive the Stokes parameters of the WS$_2$ monolayer emission (see Supplementary Materials Section II). The coherence of the WS$_2$ emission \mbox{ $\Delta = \sqrt{S_1^2+S_2^2+S_3^2} / S_0^2$} is used as the valley coherence. We find that of the measured Stokes parameters, S$_2$ and S$_3$ are negligible ($<$ 0.05, see Supplementary Materials Fig.S2), so the WS$_2$ emission does not have components that are circularly polarized or linearly polarized along the diagonal axis with respect to the excitation polarization direction.  
We calculate the valley coherence $\Delta$ at every position on the monolayer flakes. As S$_2$ and S$_3$ are negligible, the main contribution to the valley coherence is S$_1$. We confirm a non-zero WS$_2$ valley coherence, ranging from 0.1 to 0.5 depending on position and temperature. As a comparison, we perform linear polarization measurements on fl.1 and confirm that the extracted S$_1$ parameter and valley coherence are in good agreement for both measurement methods (see Supplementary Materials Fig.S2). The presence of phase coherence of the WS$_2$ emission conclusively proves that the WS$_2$ valleys indeed emit while still in a partial superposition state. 

Figures \ref{coherence}d-e present the valley coherence from (the central region of) fl.1 and fl.2 at \SI{4}{K}. As was the case for the valley polarization, the valley coherence is also significantly higher at the medians of fl.1 and the central region of fl.2 (yellow in Fig.\ref{coherence}d,e), e.g., the regions where the PL intensity is lower (compare Fig.\ref{PLpattern}e-f). The valley coherence is lower outside of the medians of fl.1 and on the edges of fl.2 (orange-red), e.g., the regions where the PL intensity is higher. The small magnitudes of S$_2$ and S$_3$ do not exhibit any position dependence (see Supplementary Materials Fig.S2). We conclude that for some flakes, the valley coherence follows the same trend as the valley polarization, being lower at high intensity regions. It is interesting to note however that fl.3 and fl.4 (see Supplementary Materials Fig.S5), even though the intensity of fl.3 is lower and the valley polarization is higher than for fl.4, do exhibit a similar valley coherence. We deduce that unlike the behaviour of the valley polarization, the inverse correlation between PL intensity and valley coherence found within a single flake is not present when comparing different WS$_2$ monolayer flakes. 

\mbox{Figure \ref{coherence}f} depicts the temperature dependence of the average valley coherence of fl.1 (open circles/stars) and fl.2 (filled circles/stars), where the values from the high and low valley coherence regions are yellow circles and red stars respectively. As was the case for the degree of valley polarization, the spatial variation in high valley coherence (yellow in Fig.\ref{coherence}d,e) and low valley coherence (red in Fig.\ref{coherence}d,e) is present at all temperatures for the same regions on the flakes: the medians on fl.1 and the central region on fl.2 (see Supplementary Materials Fig.S4 for valley coherence maps at different temperatures). Following Zhu et al, we attribute valley decoherence to the excitation of phonons and fit the temperature-dependent valley coherence with a Boltzmann distribution \cite{Zhu_WS2bilayerValleyPolarization_PNAS_2014}. If both the valley decoherence and the valley depolarization are phonon-related, this raises the question how the two processes are related. 

\subsection{Discussion}

There are various processes that lead to valley decoherence. After the excitation of both the WS$_2$ valleys ($\gamma_{excite}$ in Fig.\ref{scheme_valleys}), a number of processes may occur: either one of the electrons can hop to the other valley ($\gamma_{hopping}$), or one of the electrons can decay radiatively ($\gamma_{rad,K}$) or non-radiatevely ($\gamma_{nr,K}$). The only way for the emitted photons to be in a coherent superposition, e.g. the emitted polarization to be linearly polarized, is when both electrons emit radiatively at the same time with a fixed phase relation between them. With the observed negligible S$_2$ parameter, we can further conclude that the phase relation between the valleys is preserved, leading to the same linear polarization as the emitted light. The coherent superposition of the electrons in the two valleys will both be destroyed by valley hopping of either one of the electrons, the non-radiative decay of either one of the electrons, or when the phase relation between the two electrons is broken by further decoherence mechanisms. Thus, there are more potential processes to reduce the valley coherence than the valley polarization.  

As valley hopping is both a source of valley decoherence and of valley polarization, a finite valley polarization gives an upper limit for the valley coherence. From the temperature dependence we have deduced that both valley depolarization and valley decoherence have a phononic nature. Therefore there are two possible mechanisms to reduce the valley coherence: either phonons cause valley hopping, e.g., the same mechanism that reduces valley polarization, or phonon scattering breaks the phase relation between the electrons in the two WS$_2$ valleys. 
We observe an inverse relationship between PL intensity and valley coherence. At regions with low PL intensity we detect a high valley coherence. Therefore in these regions, the non-radiative decay rate is relatively fast with respect to the rest of the flake, and either the valley hopping rate is relatively slow, or the valleys are very well protected against phase jumps. At regions with high PL intensity we observe a low valley coherence. We conclude that in these regions, the non-radiative decay rate is relatively slow and either the valley hopping rate is relatively fast, or the phase relation between the electrons in the two valleys is easily broken. 

Thus we reported both inter-flake and intra-flake correlations between intensity, degree of valley polarization and valley coherence. It is interesting to note that the flakes that exhibited a more homogeneously distributed valley polarization and valley coherence over their surface area, do not possess a defect region (see Supplementary Materials Fig.S1). This hints at the physical importance of this defect region. A higher amount of phonons, causing valley hopping and breaking the phase relation between the WS$_2$ valleys, can come from the presence of strain, for instance caused by defects. Attributing the position-dependent variations of the PL intensity of CVD grown monolayer flakes to differences in defect density and strain is in line with previous work \cite{Gutierrez_PLpattern_NanoLett_2013, Peimyoo_growthPL_ACSNano_2013, Cong_growthPL_AOM_2020, Kim_PLpatternBiexciton_ACSNano_2016, Liu_PLpattern_NanoLetters_2016, Feng_PLpatternStrain_ACS_2017, Creary_PLpatternValley_ACSNano_2017, Rosenberger_PLpatternDefect_ACSNano_2018}. Nevertheless, the explanation of the inverse relationship between PL intensity and valley polarization and valley coherence by phonons is not straighforward. We have attributed valley depolarization and valley decoherence to phonon-induced valley hopping and phase jumps. However, an increase in the non-radiative decay is also commonly phononic in nature. If phonons both induce valley hopping lowering the valley polarization, and induce non-radiative decay that lower the intensity, phonons cannot explain the correlation between high valley polarization and low intensity. This makes the observed high valley polarization and valley coherence at regions of low PL intensity surprising. As defects and strain not only result in a larger amount of available phonons, but strain also slightly modifies the bandgap, this naturally affects the photoluminescence intensity \cite{Conley_strainBandgap_NanoLett_2013}. Therefore the observed inverse correlation between PL intensity and valley polarization and valley coherence might be related to strain. 

\section{Conclusion}

We have characterized the position-dependent polarization properties of the light emission of CVD grown WS$_2$ monolayer flakes, where the measured valley polarization sheds light on the processes within one WS$_2$ valley, and the measured valley coherence on the interaction and phase relation between both valleys. Using a rotating quarter-wave plate method to derive the Stokes parameters of the emitted light, we have characterized the full polarization state of the WS$_2$ emission and proven unambiguously the existence of coherence between the two WS$_2$ valleys. Hereby we have confirmed that the attribution of linear polarization contrast to valley quantum coherence is valid \cite{Zhu_WS2bilayerValleyPolarization_PNAS_2014, Ye_valleyCoherenceTime_WSe2_NatPhys_2017, Hao_valleyCoherenceTime_NatPhys_2016, Jones_TMDCvalleycoherence_NatNano_2013}. Moreover, we observe that both the intensity, the valley coherence and the valley polarization is non-uniform along the WS$_2$ monolayer flakes. A clear correlation exists between intensity and valley polarization, and between intensity and valley coherence, implying a correlation between radiative and non-radiative decay rates on one hand, and valley hopping and valley phase jumps on the other hand. As the temperature dependence of the valley coherence and valley polarization can be fitted with a Boltzmann distribution, we conclude that the source of valley depolarization and decoherence is phonon-related. We observe that the central regions of some of the WS$_2$ monolayer flakes exhibit a defect-related spectral behaviour. We conclude that the presence of defects, in 2D materials naturally causing strain, influences the amount of available phonons and reduces the valley polarization and the valley coherence. By investigating the full polarization-resolved emission of WS$_2$ monolayer flakes, we take the first steps in unraveling the interaction of the WS$_2$ valleys and confirming the existence of valley coherence, and thereby open the way for applications involving quantum valley manipulation and valleytronics. 

\section{Methods}

The WS$_2$ monolayer flakes are directly grown on a microchip using chemical vapour deposition (CVD) techniques. The sample preparation method is described in \cite{Irina_pyramids_2020}. 

The optical measurements are performed using a home-built spectroscopy set-up as depicted in Fig.\ref{valleyPol}b and Fig.\ref{coherence}a. The sample is placed on a piezo stage in a Montana cryostation S100, and  cooled down from room temperature to \SI{200}{K}, \SI{100}{K} and \SI{4}{K}. The sample is illuminated through an \SI{0.85}{NA} Zeiss 100x objective. All depicted polarization measurements, both the valley polarization and the valley coherence, are performed using a continuous wave laser with a wavelength of \SI{595}{nm} and a power of \SI{1.6}{mW/mm^2} (Coherent OBIS LS 594-60). The excitation light is filtered out using colour filters (Semrock NF03-594E-25 and FF01-593/LP-25). For Fig.S1a,l in the Supplementary Materials, a continuous wave laser with a wavelength of \SI{561}{nm} and a power of \SI{3.6}{mW/mm^2} is used (Cobolt 08-01/561). To avoid the depolarization consequences of tight focusing on (circular) polarization, a \SI{2}{mm} laser diameter is used, slightly underfilling the objective with a back-aperture diameter of \SI{2.8}{mm} in the excitation path. The sample emission is collected in reflection through the same objective as in excitation, and projected onto a CCD camera (Princeton Instruments ProEM 1024BX3) and spectrometer (Princeton Instruments SP2358) via a 4f lens system. 

\clearpage

\section*{Supplementary Materials}

\begin{figure*}[htp]
\centering
\includegraphics[width = 0.9\linewidth] {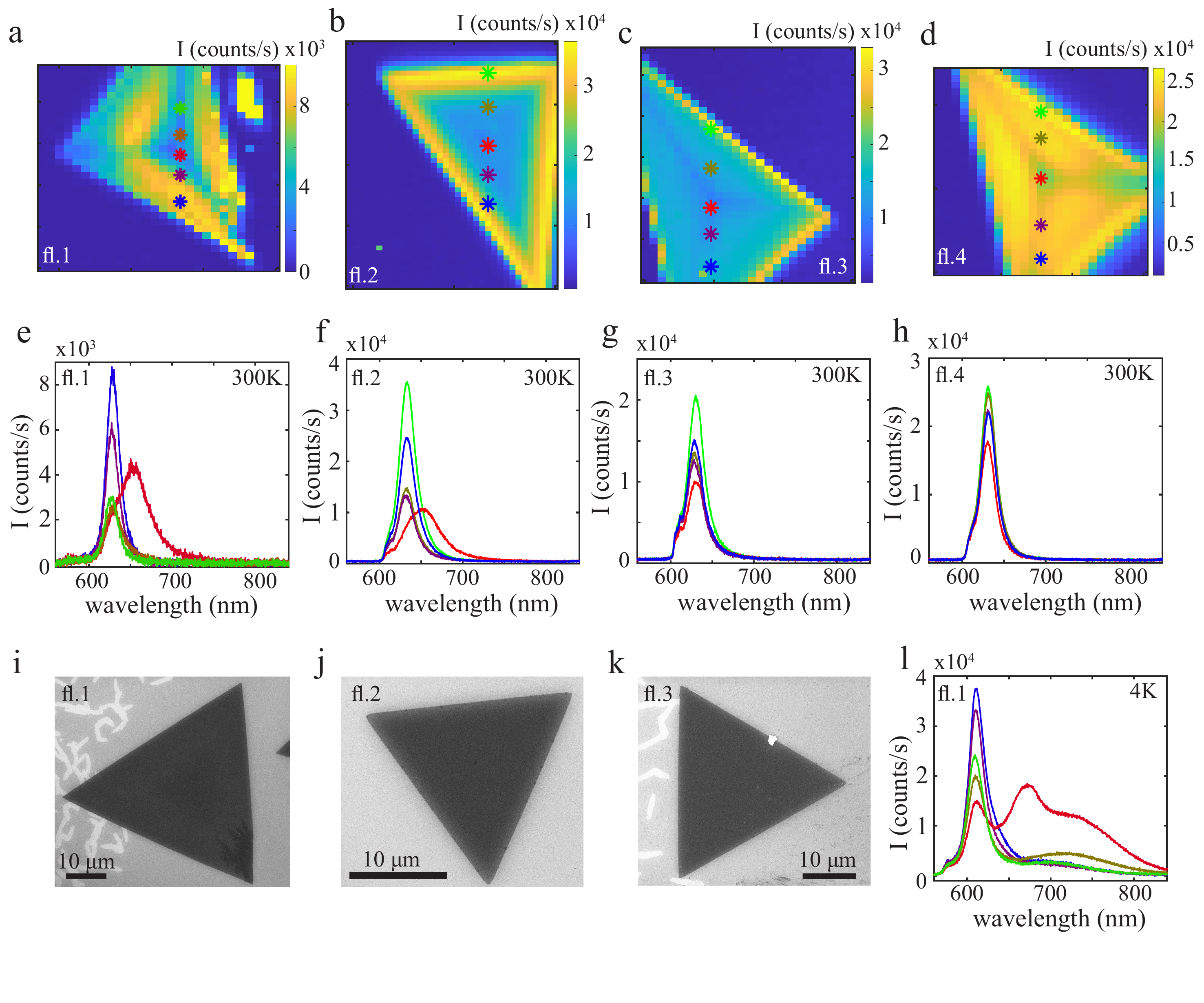}
\caption{\textbf{Defect peak on monolayer flakes} \\
\textbf{a-d.} Raster scan of the PL intensity over fl.1 - fl.4, acquired at room temperature. Note that the x and y axis in the raster scan are slightly skewed due to experimental constraints (see Section \ref{sect_experimental}) The stars indicate the positions of \textbf{e-h.} the PL spectral response taken around the flake centre. For fl.1 (\textbf{e.}), the spectral peaks on the medians (green and brown) have a lower intensity but the same shape as the spectra on the flake (purple and blue). The same holds true for fl.2 (\textbf{f.}), where the spectral response on the edge (green and blue) has a higher intensity as the spectral response in the middle (brown and purple). The spectral peaks in the middle of fl.1 and fl.2 (in red) are significantly red shifted. \textbf{g-h.} The spectra of fl.3 and fl.4 do not reveal defect-like behaviour. The only difference between spectra on different position of these flakes, is the intensity. \textbf{i-k.} SEM images of fl.1 - fl.3 do not reveal any grain boundaries or other structural reasons that would explain the non-uniformity in the PL intensity. \textbf{l.} PL spectra of fl.1 taken at \SI{4}{K}. The spectrum in the very middle of the flake (in red) consists of at least three peaks, one at the exciton position and two red shifted to higher wavelengths.}
\label{defect_other_flakes}
\end{figure*}

\subsection{Spectral behaviour of monolayer flakes}

The WS$_2$ monolayer flakes fl.1 and fl.2 exhibit a significant spectral difference when excited in the central region with respect to the rest of the flakes. \mbox{Figures \ref{defect_other_flakes}a-b} depict the PL intensity of fl.1 and fl.2 at room temperature. \mbox{Figures \ref{defect_other_flakes}e,l} present spectra (of fl.1) at the flake medians (green and brown), the middle (red) and on the rest of the monolayer (purple and blue) (positions are indicated by stars in Fig.\ref{defect_other_flakes}a). The spectra at the medians and on the rest of the monolayer are not fundamentally different, except for the overall photoluminescence intensity. The spectral response in the middle of the monolayer fl.1 is red shifted towards \SI{650}{nm} at room temperature (red spectrum in Fig.\ref{defect_other_flakes}e) and consists of at least three peaks at \SI{4}{K}, one at the exciton position and two with lower energies (red spectrum in Fig.\ref{defect_other_flakes}l). The spectra of fl.2 in Fig.\ref{defect_other_flakes}f exhibit a similar behaviour as of fl.1. The spectra on the bright edges of the flake (blue and green) and in the low intensity region (purple and brown) only differ in intensity, but the spectrum in the middle (in red) is red shifted. 

Figure \ref{defect_other_flakes}c-d present the PL intensity of fl.3 and fl.4 at room temperature (fl.3 is \SI{35}{\mu m} and fl.4 is \SI{15}{\mu m} large). The PL intensity is not homogeneous over these two flakes, but is lower at the triangle medians (like fl.1). The PL intensity of the edges of fl.3 is higher (like fl.2). However, the PL spectra of fl.3 and fl.4 in Fig.\ref{defect_other_flakes}g,h all have the same shape and only differ in intensity. Even the spectra in the very middle of these flakes (in red) do not exhibit a red-shifted peak as the spectra of fl.1 and fl.2. 

Possible explanations for the red shifted spectral response in the middle of the monolayer flakes are either trion or defect related. The trion peak is expected to be close to the spectral position of the exciton (within \SI{10}{nm}) \cite{Plechinger_WS2trionsTemp_PSS_2015, Currie_WS2trion_APL_2015, Kato_WS2trions_ACSNano_2016, Krustok_WS2defect_AIP_2017, Jadczak_TMDtrionsTemp_Nanotech_2017}, whereas some reported WS$_2$ defect peaks are further red-shifted from the exciton (to around \SI{650}{nm}) \cite{Currie_WS2trion_APL_2015, He_WS2defect_ACSNano_2016}. Spectra measured in different studies vary greatly, indicating large and possibly unknown sample differences. Based on the spectral position we conclude that the red-shifted spectral response in the very middle of fl.1 and fl.2 might be induced by the presence of local structural defects rather than being the trion. 

Scanning electron microscopy (SEM) images of fl.1 - fl.3, presented in Fig.\ref{defect_other_flakes}i-k, do not reveal any grain boundaries or other structural reasons that would explain the non-uniformity in the PL intensity. Structural characterization on the atomic level cannot be performed on the investigated flakes, as the relatively thick silicon substrate is not transparent to Transmission Electron Microscopy.  

\subsection{Stokes analysis}

As mentioned in the main text, we determine the valley coherence using the rotating quarter-wave plate method (see Fig.4a in the main text). Given a quarter-wave plate (QWP) with perfect retardance, the measured intensity can be written as a function of the Stokes parameters and the QWP angle using Mueller matrix multiplication as \cite{Schaefer_stokesPol_APL_2007, collett_quarterwaveplate_1993, collett_quarterwaveplate_2005}:

\begin{equation}
    I(\theta) = \frac{1}{2} (S_0 + S_1 cos^2 2\theta + S_2 cos2\theta sin2\theta + S_3 sin2\theta ) ,
\end{equation}

\noindent with $\theta$ the angle of the QWP and S$_0$ to S$_3$ the Stokes parameters. This can be rewritten to a truncated Fourier series, after which the Stokes parameters can be found using Fourier analysis \cite{Schaefer_stokesPol_APL_2007, collett_quarterwaveplate_1993, collett_quarterwaveplate_2005}. Unfortunately, the experimentally more likely situation of a QWP with non-perfect retardance, as we use in our experiment, is less straightforward (see Section III for more details about the used QWP). In that case, the measured intensity is:

\begin{equation}
    \begin{split}
        I(\theta,\delta) = \frac{1}{2} (S_0 + S_1 (cos^2 2\theta + sin^2 2\theta cos \delta) + \\
        + S_2 (cos 2\theta sin 2\theta)(1-cos \delta) + S_3 sin 2\theta sin \delta ,
    \end{split}
    \label{equation_mueller}
\end{equation}

\begin{figure*}[htp]
\centering
\includegraphics[width = 0.9\linewidth] {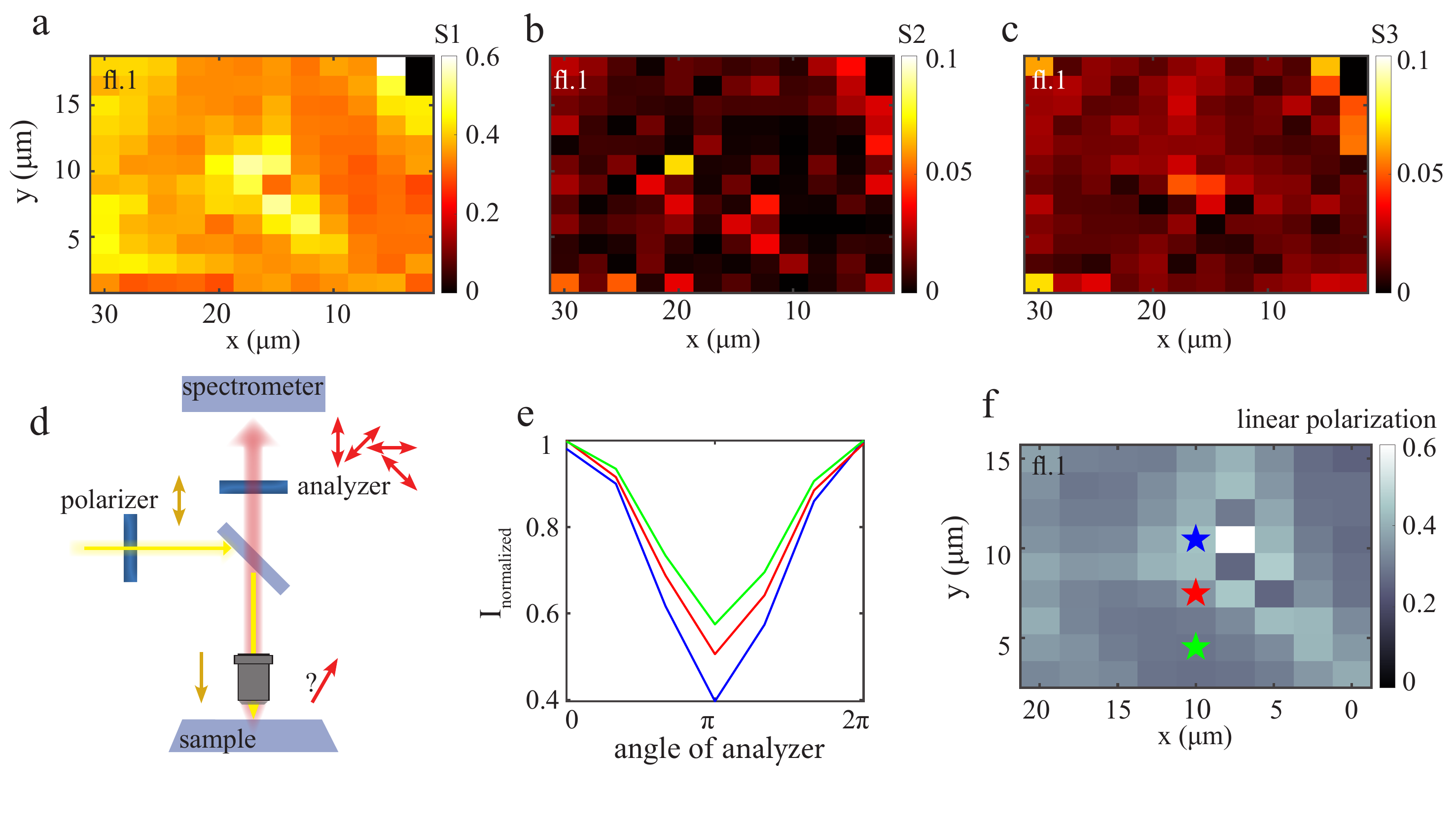}
\caption{\textbf{Stokes analysis and linear polarization contrast} \\
\textbf{a-c} Maps of the Stokes parameters of the PL of fl.1, derived using the rotating quarter-wave plate method (see main text). It is clear in \textbf{b,c} that S$_2$ and S$_3$ are negligible and independent of position. Therefore the main contribution to the valley coherence of the emission from the WS$_2$ valleys is S$_1$. In \textbf{a}, the S$_1$ parameter of the WS$_2$ emission has higher values at low intensity regions of fl.1, e.g. the triangle medians, and lower values at the other flake regions. The S$_1$ values are comparable to the valley coherence (compare with Fig.4d in the main text). \textbf{d} We also perform linear polarization contrast measurements as performed by other groups \cite{Zhu_WS2bilayerValleyPolarization_PNAS_2014, Kim_linearPolMoS2_KoreanPhys_2015, Ye_valleyCoherenceTime_WSe2_NatPhys_2017, Hao_valleyCoherenceTime_NatPhys_2016}. The set-up is comparable to Fig.4a in the main text: we excite with linearly polarized light, but now we analyze the emission polarization using a rotating analyzer. \textbf{e} Normalized intensity of the PL emission from fl.1 at different positions, as a function of the analyzer angle. From these curves, we determine the degree of linear polarization by $\frac{I_0 - I_\pi}{I_0 + I_\pi}$. \textbf{f} Map of the derived degree of linear polarization of the PL of fl.1 at \SI{4}{K} (stars indicate the position of the curves in \textbf{e}). Note that the degree of linear polarization is very similar to the S$_1$ and to the valley coherence, which confirms that measuring the degree of linear polarization of WS$_2$ emission is a valid method for determining the valley coherence. }
\label{Stokes_linear}
\end{figure*}

\noindent with $\delta$ the retardance of the QWP. Fig.4c in the main text depicts the measured PL intensity for different angles of the QWP, where the fit of the curve is based on Eq.\ref{equation_mueller}. Figure \ref{Stokes_linear}a-c depicts the Stokes parameters as a result of the fitted PL intensity of fl.1 at \SI{4}{K}. When inspecting Fig.\ref{Stokes_linear}b and c, it becomes apparent that the S$_2$ and S$_3$ parameters of the WS$_2$ emission (upon excitation with linearly polarized light) are negligible and independent of position. The only relevant Stokes parameter is S$_1$. Therefore, comparing the calculated valley coherence $\Delta = \sqrt{S_1^2+S_2^2+S_3^2} / S_0$ of fl.1 in Fig.4d in the main text to the S$_1$ in Fig.\ref{Stokes_linear}a, it is clear that the value and the position dependence is very similar: S$_1$ is also higher along the flake medians than on the rest of fl.1, with values between 0.4 and 0.6. This indicates that the previous measurements of linear polarization contrast \cite{Zhu_WS2bilayerValleyPolarization_PNAS_2014, Kim_linearPolMoS2_KoreanPhys_2015, Ye_valleyCoherenceTime_WSe2_NatPhys_2017, Hao_valleyCoherenceTime_NatPhys_2016} did indeed point at valley coherence. 

To confirm these conclusions and compare the results to the rotating QWP experiment, we perform a linear polarization measurement on fl.1. Figure \ref{Stokes_linear}d depicts the used set-up, where the sample is excited with linearly polarized light, and the resulting emission is analyzed by a polarization analyzer. Figure \ref{Stokes_linear}e depicts the WS$_2$ PL intensity as a function of the angle of the polarization analyzer, for different positions on the flake. The dip at $\pi$ is at most 0.40. We calculate the linear polarization contrast $\frac{I_0 - I_\pi}{I_0 + I_\pi}$ and plot this for fl.1 in Fig.\ref{Stokes_linear}f (the stars indicate the position of the curves in Fig.\ref{Stokes_linear}e). Comparing S$_1$ in Fig.\ref{Stokes_linear}a, determined using the rotating QWP method, and the linear polarization contrast in Fig.\ref{Stokes_linear}f, we conclude that the values and the pattern of high S$_1$ at the flake medians and low S$_1$ at the rest of the flake, is present in both experimental methods. Therefore we confirm that the WS$_2$ emission does indeed exhibit a coherence, indicating a partial quantum entanglement of the WS$_2$ valleys. 

\subsection{Experimental methods} \label{sect_experimental}

\begin{figure*}[htp]
\centering
\includegraphics[width = 0.9\linewidth] {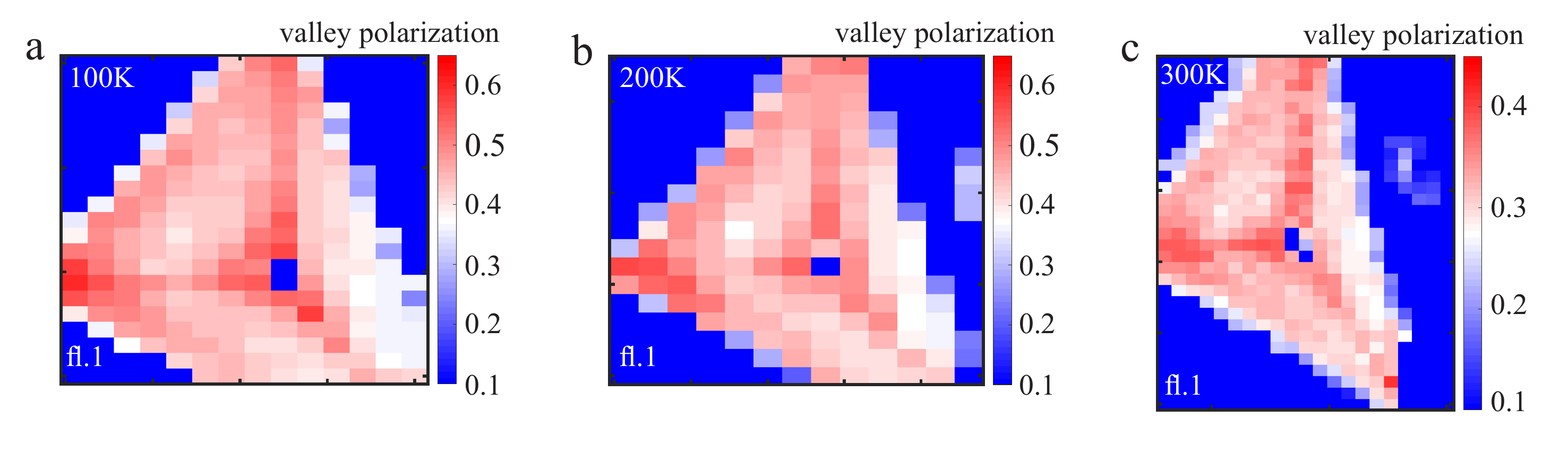}
\caption{\textbf{Temperature-dependent valley polarization} \\
\textbf{a-c} Valley polarization of fl.1 at 100 - \SI{300}{K}. Just like the valley polarization at \SI{4}{K} (compare Fig.2d in the main text), the regions of the flake with lower PL intensity, e.g. the triangle medians, exhibit higher valley polarization than the higher intensity regions on this monolayer flake. This valley-polarization pattern persists from \SI{4}{K} to room temperature. Note that we do not calculate the valley polarization of spectra of the defect region, in the middle of the flake.}
\label{valleyPol_temperature}
\end{figure*}

\begin{figure*}[htp]
\centering
\includegraphics[width = 0.9\linewidth] {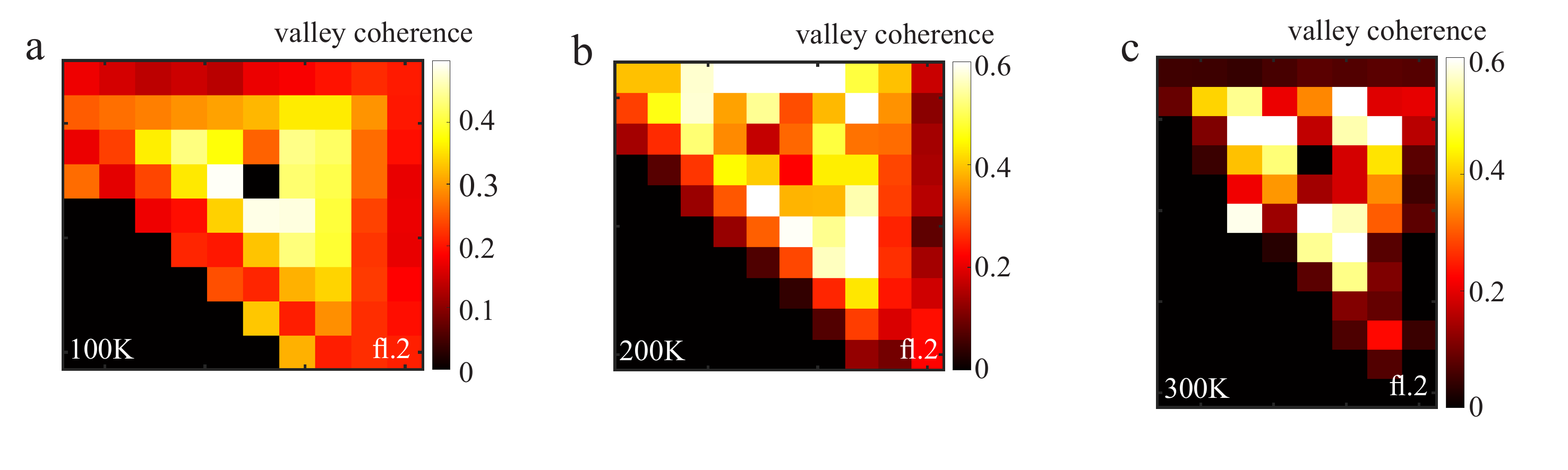}
\caption{\textbf{Temperature-dependent valley coherence} \\
\textbf{a-c} Valley coherence of fl.2 at 100 - \SI{300}{K}. Just like the valley coherence at \SI{4}{K} (compare Fig.4e in the main text), the middle region of the flake with lower PL intensity exhibits higher valley coherence than the higher intensity edges on this monolayer flake. The difference in higher and lower valley coherence is present even at room temperature, although not as clear as at cryogenic temperatures. Note that we do not calculate the valley coherence of the defect-related spectral response in the middle of the flake.}
\label{valleyCoherence_temperature}
\end{figure*}

As mentioned in the main text, the sample is placed on a piezo stage in a Montana cryostation S100. The raster scans presented in this work are performed using Attocube ANPxy101/RES piezo scanners. There is a small cross-coupling between the in-plane and out-of-plane piezo scanners, causing a skewing between the x and y axis of the raster scans depicted in this work. Comparing the raster scans with the optical and SEM images of the monolayer flakes, we can nevertheless correlate the position of certain spectral features with the position on the flake.

As mentioned in the main text, the optical measurements are performed using a home-built spectroscopy set-up. For the rotating quarter-wave plate (QWP) experiments we use a zero-order QWP for \SI{633}{nm} (Thorlabs WPQ05M-633). The retardance of this waveplate depends on the wavelength of the incident light, and ranges from $\delta$=0.25 for \SI{633}{nm} to $\delta$=0.265 for \SI{600}{nm}. For cryogenic temperatures, the bandgap energy of WS$_2$ red shifts to \SI{605}{nm}. Therefore, this zero-order QWP works less well for WS$_2$ emission at cryogenic temperatures then at room temperature. We have taken this wavelength-dependence into account when fitting the angle-dependent PL intensity in Fig.4c in the main text. 

Even though a superachromatic waveplate would have had a retardance close to 0.25 for the range of the emission wavelengths, it has an experimentally much larger disadvantage. Most superachromatic waveplates cause a beam deviation up to \SI{3}{arcmin} due to the necessary composition of at least three separate birefringent plates, whereas the beam deviation caused by a zero-order waveplate is an order of magnitude lower, up to \SI{10}{arcsec}. That means that even with a perfect alignment, light that passed through the superachromatic waveplate will travel under a different angle, depending on the waveplate rotation. As the emission travels to the CCD camera through a spectrometer, angle differences in the incident light will propagate and cause an unwanted dependence of the measured intensity on the waveplate angle. Therefore we have chosen to use the zero-order QWP, taking into account its wavelength-dependent retardance. We attribute the remaining difference in intensity at $\pi/4$ and $3\pi/4$ in Fig.4c of the main text with respect to the fitted curve, to small angular deviations caused by the alignment of the zero-order waveplate and the emission path. 

\begin{figure*}[htp]
\centering
\includegraphics[width = 0.9\linewidth] {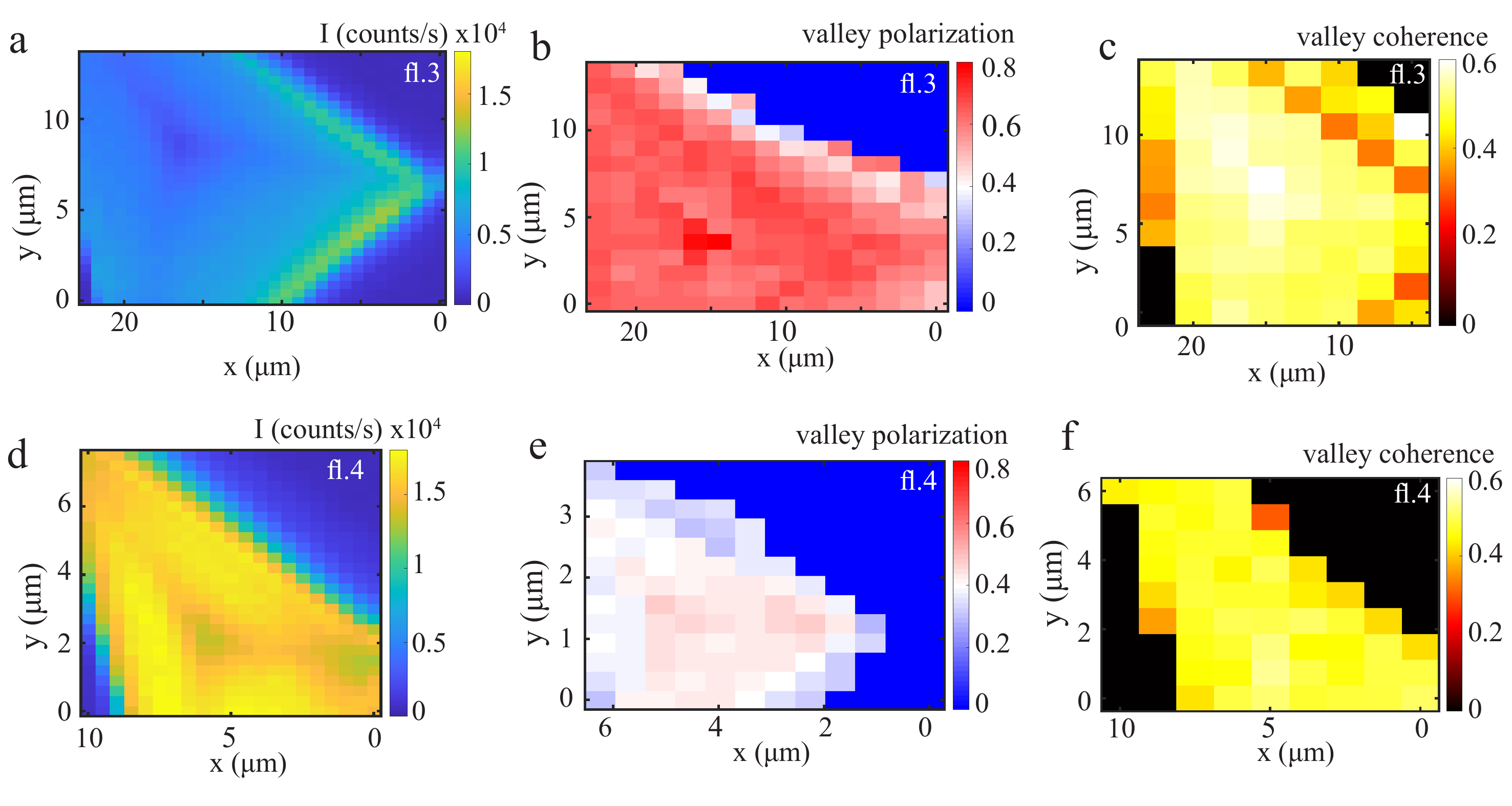}
\caption{\textbf{PL intensity, valley polarization and valley coherence of fl.3 and fl.4} \\
\textbf{a,d} PL intensity of fl.3 and fl.4 at \SI{4}{K}, representing the monolayer flakes that do not exhibit a defect region in the middle. The extent of the colormaps is chosen the same for easy comparison. As could be seen in Fig.\ref{defect_other_flakes}c-d, the intensity of fl.3 is lower than fl.4. \textbf{b,e} Valley polarization of fl.3 and fl.4 at \SI{4}{K}. Note that the valley polarization of these flakes is much more homogeneous than of fl.1 and fl.2. Fl.3 has a lower PL intensity, and a higher valley polarization, when compared to fl.4. This inter-flake inverse relationship between intensity and valley polarization follows the same trend as the intra-flake relationship seen in fl.1 and fl.2. \textbf{c,f} Valley coherence of fl.3 and fl.4 at \SI{4}{K}. Note that now the valley coherence of these flakes is comparable, although fl.3 has slightly higher values. This means that the inverse relationship between PL intensity and valley coherence for regions in fl.1 and fl.2, cannot be confirmed within the experimental margins of error for fl.3 and fl.4. }
\label{otherFlakes}
\end{figure*}

\subsection{Temperature dependence}

As mentioned in the main text, the valley polarization on the monolayer flakes follows the same distribution as the intensity. Figure \ref{valleyPol_temperature}a-c present the valley polarization of fl.1 at \SI{100}{K}, \SI{200}{K} and room temperature. The resulting pattern of high valley polarization at the flake medians and lower valley polarization at the rest of the flake is the same as the pattern at \SI{4}{K} in Fig.2d in the main text. The pattern in the PL intensity also exists at all temperatures, compare Fig.\ref{defect_other_flakes}c,d and Fig.\ref{otherFlakes}a,d. We conclude that the inverse correlation between valley polarization and PL intensity exists at all temperatures. 

The same holds for the valley coherence. Figure \ref{valleyCoherence_temperature}a-c present the valley coherence of fl.2 at \SI{100}{K}, \SI{200}{K} and room temperature. The pattern of high valley coherence in the middle region and low valley coherence on the edges is the same as the pattern at \SI{4}{K} in Fig.4e in the main text, although the pattern is not as clear at room temperature. We conclude that the inverse correlation between valley coherence and PL intensity also exists at all temperature. 

\subsection{Polarization behaviour of flake 3 and flake 4}

Figure \ref{otherFlakes}a,d presents the PL intensity of fl.3 and fl.4 at \SI{4}{K}. Where fl.1 and fl.2 exhibited a defect-related region in the middle, fl.3 and fl.4 do not (see Fig.\ref{defect_other_flakes}). The PL intensity of fl.3 is lower than that of fl.4 (the scale bars are chosen the same for direct comparison). Figure \ref{otherFlakes}b,e presents the valley polarization of fl.3 and fl.4 at \SI{4}{K}. The valley polarization of these monolayer flakes is more homogeneous over the flakes. We hypothesise that this is related to the absence of the defect region in these flakes. The valley polarization of fl.3 is clearly larger than that of fl.4, therefore following the same inverse correlation between PL intensity and valley polarization as the different regions on fl.1 and fl.2. 

However, the inverse correlation between PL intensity and valley coherence is not as clear for these flakes. Figure \ref{otherFlakes}c,f present the valley coherence of fl.3 and fl.4. The valley coherence of fl.3 is only slightly higher than that of fl.4 (0.55 vs. 0.45), but the values are of the same order of magnitude.

\bibliography{article_polarization_flakes}

\end{document}